\documentclass{elsart}
\usepackage{amsmath,bm}
\usepackage{graphicx}

\usepackage[normalem]{ulem}  
\usepackage[dvips]{color} 

\begin{document}

\begin{frontmatter}

\title{Nuclear collision in strong magnetic field}

\author{Gao-Chan Yong}

\address{Institute of Modern Physics, Chinese Academy of
Sciences, Lanzhou 730000, China}

\begin{abstract}
Based on the Boltzmann-Uehling-Uhlenbeck transport model coupled
with the Lorentz force equation, we studied nucleus-nucleus
collision in strong magnetic field. We find that neutrons and
protons can be separated from a nucleus by strong magnetic field
and neutron-rich high density nuclear matter and low density
proton collectivity matter can be formed during nucleus-nucleus
collision. The electric field produced by proton collectivity can
accelerate proton and charged meson up to very high energies.
Besides the studies of isospin physics such as symmetry energy,
these results may help us to understand the acceleration
mechanisms of high energy charged particles in the cosmic rays.
\end{abstract}

\begin{keyword}
Nuclear collision, strong magnetic field, isospin physics,
acceleration mechanisms, origin of cosmic rays.
 \PACS 25.70.-z \sep 24.10.Lx \sep 96.50.S-
\end{keyword}

\end{frontmatter}

The question of the origin of cosmic rays continues to be regarded
as an unsolved problem even after almost one century years of
research since the announcement of their discovery in 1912
\cite{Bha00}. The cosmic rays of extremely high-energy pose a
serious challenge for conventional theories of origin of cosmic
rays based on acceleration of charged particles in powerful
astrophysical objects. The question of origin of these extremely
high-energy cosmic rays is currently a subject of much intense
debate and discussions \cite{Cro99}. On the origin of energetic
cosmic-rays in the universe, except through decay of sufficiently
massive particles originating from processes in the early
universe, there are basically two kinds of acceleration mechanisms
considered in connection with cosmic rays acceleration, i.e.,
direct acceleration of charged particles by an electric field and
statistical acceleration in a magnetized plasma. In the direct
acceleration mechanism, the electric field in question is
generally due to a rotating magnetic neutron star (pulsar) or, a
(rotating) accretion disk threaded by magnetic fields, etc.
\cite{Bha00}. In this article, we show that nuclear collision in
strong magnetic field can produce proton collectivity in space,
protons or other charged particles thus can be accelerated
directly by electric fields, which are produced by positive
charges, not by changing magnetic fields.

The condition of strong magnetic field may exist in the universe,
such as white dwarfs, neutron stars, and accretion disks around
black holes, and the maximum value of magnetic fields in the
universe may reach $10^{20}\sim 10^{42}$ G \cite{sha06}. And with
the rapid development of laser technology, obtaining strong
magnetic field artificially in terrestrial laboratory also may be
possible \cite{led03,um00}. Although nuclear collisions exist
extensively in the universe, up to now, nuclear reactions in the
strong magnetic field were seldom reported. The separation of
neutrons and protons from a nucleus while nucleus-nucleus
collision in strong magnetic field may shed light on the
acceleration mechanisms of high energy charged particles in the
cosmic rays.

Nuclear matter studies can be carried out semiclassically
following the so called nuclear pasta phases
\cite{sp08,hor04,to05,sp10}. In this study we also use the
semiclassical isospin-dependent Boltzmann-Uehling-Uhlenbeck (BUU)
transport model \cite{bertsch88,baoan97}, which is quite
successful in describing dynamical evolution of nuclear collision.
The BUU equation describes time evolution of the single particle
phase space distribution function $f(r,p,t)$, it reads:
\begin{equation}
\frac{\partial f}{\partial t}+\upsilon\cdot\nabla_{r}f-\nabla
U\cdot\nabla_{p}f=I_{collision}. \label{BUU}
\end{equation}
$f (r,p,t)$ can be viewed semi-classically as the probability of
finding a particle at time $t$ with momentum $p$ at position $r$.
The mean-field potential $U$ depends on position and momentum of
the particle and is computed self-consistently using the
distribution functions $f (r,p,t)$. The collision item
$I_{collision}$ on the right-hand side of Eq.~(\ref{BUU}) governs
the modifications of $f (r,p,t)$ by elastic and inelastic two body
collisions caused by short-range residual interactions
\cite{bertsch85}. The proton and neutron densities of colliding
nuclei are given by Skyrme- Hartree-Fock with Skyrme $M^{*}$ force
parameters \cite{Friedrich86}. The isospin dependence is included
in the dynamics through nucleon-nucleon collisions by using
isospin-dependent cross sections and Pauli blocking factors
\cite{yong06}. The isospin and momentum-dependent mean field
potential used is \cite{bcs04}
\begin{eqnarray}
U(\rho, \delta, \textbf{p},\tau)
=A_u(x)\frac{\rho_{\tau^\prime}}{\rho_0}+A_l(x)\frac{\rho_{\tau}}{\rho_0}\nonumber\\
+B\left(\frac{\rho}{\rho_0}\right)^\sigma\left(1-x\delta^2\right)\nonumber
-8x\tau\frac{B}{\sigma+1}\frac{\rho^{\sigma-1}}{\rho_0^\sigma}\delta\rho_{\tau^{\prime}}\nonumber\\
+\sum_{t=\tau,\tau^{\prime}}\frac{2C_{\tau,t}}{\rho_0}\int{d^3\textbf{p}^{\prime}\frac{f_{t}(\textbf{r},
\textbf{p}^{\prime})}{1+\left(\textbf{p}-
\textbf{p}^{\prime}\right)^2/\Lambda^2}}. \label{Un}
\end{eqnarray}
In the above equation, $\delta =(\rho _{n}-\rho _{p})/(\rho
_{n}+\rho _{p})$ is the isospin asymmetry parameter, $\rho =\rho
_{n}+\rho _{p}$ is the baryon density and $\rho _{n}, \rho _{p}$
are the neutron and proton densities, respectively. $\tau
=1/2(-1/2)$ for neutron (proton) and $\tau \neq \tau ^{\prime }$,
$\sigma =4/3$, $f_{\tau }(\mathbf{r},\mathbf{p})$ is the
phase-space distribution function at coordinate $\mathbf{r}$ and
momentum $\mathbf{p}$. The parameters $A_{u}(x),A_{l}(x),B,C_{\tau
,\tau }$, $C_{\tau ,\tau ^{\prime }}$ and $\Lambda $ were set by
reproducing the momentum-dependent potential $U(\rho ,\delta
,\mathbf{p},\tau )$ predicted by the Gogny Hartree-Fock and/or the
Brueckner-Hartree-Fock calculations \cite{das03,bom91}, the
saturation properties of symmetric nuclear matter and the symmetry
energy of about $31.6$ MeV at normal nuclear matter density $\rho _{0}=0.16$ fm%
$^{-3}$. The incompressibility of symmetric nuclear matter at
normal density is set to be $211$ MeV. The parameters $A_{u}(x)$
and $A_{l}(x)$ depend on the parameter $x$ according to
\begin{equation}
A_{u}(x)=-95.98-\frac{2B}{\sigma +1}x, A_{l}(x)=-120.57+\frac{2B}{%
\sigma +1}x,
\end{equation}%
where $B=106.35$ MeV. $\Lambda =p_{F}^{0}$ is the nucleon Fermi
momentum in
symmetric nuclear matter, $C_{\tau ,\tau ^{\prime }}=-103.4$ MeV and $%
C_{\tau ,\tau }=-11.7$ MeV. The $C_{\tau ,\tau ^{\prime }}$ and
$C_{\tau ,\tau }$ items are the momentum-dependent interactions of
a nucleon with unlike and like nucleons in the surrounding nuclear
matter. The parameter $x$ is introduced to mimic various
density-dependent symmetry energies $E_{\text{sym}}(\rho )$
predicted by microscopic and phenomenological many-body
approaches. Because we do not study nuclear symmetry energy here,
in this article we just let the variable $x$ be $1$ \cite{xiao09}.
The isoscalar part $(U_{n}+U_{p})/2$ of the single nucleon
potential was shown to be in good agreement with that of the
variational many-body calculations and the results of the BHF
approach including three-body forces. The isovector part
$(U_{n}-U_{p})/2\delta $ is consistent with the experimental Lane
potential \cite{yong06}. We use the isospin-dependent in-medium
reduced nucleon-nucleon elastic scattering cross section from the
scaling model according to nucleon effective mass. For in-medium
nucleon-nucleon inelastic scattering cross section, we at present
use the free nucleon-nucleon inelastic scattering cross section
\cite{yong09}.

Updates of nucleonic momentum are generally owing to momentum and
spatial location dependence of its mean-field potential $U$,
decided by the gradient force $\nabla U$ in Eq.~(\ref{BUU}).
Besides the gradient and Coulomb forces added on the charged
particles, momentum of charged particle also changes owing to the
Lorentz force. For the additional magnetic filed force of charged
hadron, we use the Lorentz force equation
\begin{equation}\label{lorentz}
\overrightarrow{F}=q\overrightarrow{v}\times\overrightarrow{B}.
\end{equation}
Where $\overrightarrow{v}$ is the velocity of charged particle and
$\overrightarrow{B}$ is the additional magnetic field. In the
practical calculations we add the Lorentz force to the Coulomb and
mean-field gradient forces. Because the strength of magnetic field
in this study is huge, we in the present simulations use the
relativistic form of the Lorentz force equation and let the time
step interval $dt$ of updates of particle's phase space
information be a very small value ($dt=0.0025 fm/c$).

Studying nucleus-nucleus collision in strong magnetic field, we
should firstly get a picture of how the collision evolves with or
without magnetic field. Fig.~1 shows nuclear collision process
with and without strong magnetic field. From upper panels we can
see that, without magnetic field, protons and neutrons have almost
the same mode of motions. Protons and neutrons go through the same
compressions and inflations. These are normal knowledge of nuclear
reaction. With strong magnetic field, however, the whole situation
is changed. From the lower panels we can first see that, protons
in the target and projectile refuse to collide and look like
keeping still collectively owing to the Lorentz forces added on
the protons. Whereas the neutrons in the target and projectile
trend to collide, just like the case without magnetic field. The
colliding neutrons form high density neutron matter transiently.
The ``still'' protons of the target and projectile form low
density proton matter. We dub neutron matter or proton matter
asymmetric nuclear matter. If the strength of magnetic field is
smaller than $\sim10^{17}$ Tesla, the Lorentz force can not
overcome the mean-field gradient force. The separation thus can
not happen. With such strong magnetic field, the protons almost
keep still in space, i.e., fixed in coordinate space by the
magnetic field and can be kept macroscopically. The separation of
neutrons and protons from a nucleus with strong magnetic field may
shed light on the possible origin of cosmic rays. If the magnetic
fields are not homogeneous, such as the wandering magnetic fields,
the separation of neutrons and protons within a nucleus can be
kept ultimately. And the separating neutrons and protons move
respectively. In the universe, the collision between nuclei may be
replaced by physical collisions between stars
\cite{ben87,fre04,zwart99}. Collisions among light nuclei in stars
or interstellar matter with magnetic field are similar to the
$^{124}$Sn+$^{112}$Sn collision here. In case the magnetic field
decreases or disappears, the existing proton collectivity
disperses promptly owing to the Coulomb actions.

To show the acceleration mechanisms of high energy charged
particles in the cosmic rays by the electric field, we draw the
sketch-map of Fig.~2. The left-hand side of Fig.~2 is the
sketch-map of proton collectivity forms in magnetic field via
large number of ``nucleus-nucleus collision'' (in fact it is not
nucleus-nucleus collision, but mainly is neutron-neutron
collisions and protons keep ``still''). When two dense stars with
strong magnetic field collide, as shown in Fig.~2, large number of
proton collectivity can be formed. They are fixed in certain space
by the strong magnetic field so that the coulomb potentials are
stored. While without strong magnetic field, however, this
configuration changes promptly. The right-hand side shows the case
without strong magnetic field. Protons in the collectivity
disperse in all directions with the actions of the Coulomb
potential. The first escaped protons or the boundary protons will
be greatly accelerated, so they possess large velocities. The
inner protons will have small velocities. Some escaped protons can
be accelerated again and again here and there. This may be a
reason why the cosmic-ray composition (above 1.6 EeV) is
proton-dominated \cite{fermi49,Bell99,abb2010}. The energy of
proton accelerated can be roughly calculated via
\begin{equation}\label{energy}
U=\sum_{q_{i}=1}^{q_{i}=Q}\frac{1}{4\pi\varepsilon_{0}}\frac{q_{i}}{r_{i}}\sim0.047\frac{Q}{R}~
(GeV).
\end{equation}
Here $U$ is the Coulomb energy (in unit of GeV), $Q$ (in unit of
elementary charge) and $R$ (in unit of fermi) are the accelerating
charges and mean distance between accelerated proton and
accelerating charges, respectively. If the value $Q$ is large
enough, the proton then can obtain a energy of several or even
tens/hundreds of GeV. Assuming the accelerated proton mean $R$ is
100 fm and proton density in collectivity (for proton
collectivity, we suppose a semi-spherical shell distribution) is
the same as that in $^{208}$Pb, the proton then can obtain a
maximal energy of about $0.03\times R/fm \times GeV = 3 GeV$.
Practically, scale of proton collectivity formed in strong
magnetic field may be much larger than 100 fm, the proton energy
is thus also much larger than $3 GeV$. Note here that proton may
be accelerated many times as shown on the right-hand side of
Fig.~\ref{acc} owing to the winding magnetic field. Therefore the
energetic cosmic-ray protons or other charged particles may be
accelerated by electric field. As for other cosmic-ray nuclei,
they can be accelerated directly by the electric field, having no
colliding separation of neutrons and protons via nucleus-nucleus
collision.

Because of the separation of neutrons and protons within a nucleus
while nucleus-nucleus collision in strong magnetic field, the
compressed nuclear matter must be neutron-rich matter. Fig.~3
shows neutron to proton ratio $n/p$ of such high density nuclear
matter formed during the collision with strong magnetic field.
Here the $n/p$ of high density nuclear matter formed can reach 4
(if the colliding nuclei are far apart, the $n/p$ of compressed
nuclear matter can be even more large), is much larger than
obtained from general nuclear reaction in terrestrial laboratory,
in which the $n/p$ of transiently formed nuclear matter is about
1. If in the universe, somewhere there are strong magnetic fields
as large as $10^{17}$ Tesla, one then can possibly in heaven use
such asymmetric nuclear matter to study isospin physics (physics
relevant to unequal numbers of neutrons and protons), such as
nuclear matter symmetry energy (energy relevant to the changing of
mean energy per nucleon owing to unequal numbers of protons and
neutrons, is a main subject of isospin physics), which is crucial
for understanding many interesting issues in both nuclear physics
and astrophysics
\cite{Bro00,Dan02a,Bar05,LCK08,Sum94,Lat04,Ste05a} and has been
regarded as the most uncertain property of dense neutron-rich
nuclear matter \cite{Kut94,Kub99}. The advantage of using such
asymmetric nuclear matter to study isospin physics is that such
asymmetric nuclear matter formed in strong magnetic field via
nuclear collision has huge asymmetry, effects of isospin are thus
enlarged remarkably. Ever since a long time ago, the small
asymmetry of compressed nuclear matter obtained in terrestrial
laboratory has been one of the most troublesome factors (which is
owing to small resolution of isospin effect caused by small
asymmetry of nuclear matter produced through ordinary nuclear
collision) to study isospin physics, especially for nuclear
symmetry energy.

Pion production in nuclear reaction nowadays has attracted much
attention in nuclear physics community \cite{Bar05,LCK08,yong09}.
One important reason is that pion production is connected with the
high density behavior of nuclear symmetry energy \cite{LiBA02}.
Here we only study pion meson in original cosmic rays. Fig.~4
shows evolution of $\pi^-/\pi^+$ ratio with and without magnetic
field, using $^{124}$Sn+$^{112}$Sn collision at 400 MeV/nucleon as
a example. At early stage of nucleus-nucleus collision in the
magnetic field, as shown in Fig.~\ref{asy}, we can see that
neutron-neutron collisions can happen while protons nearly do not
take part in the collisions. We thus see a very large
$\pi^-/\pi^+$ ratio with magnetic field ($\pi^-$'s are mainly from
neutron-neutron collisions, $\pi^+$'s are mainly from
proton-proton collisions, neutron-proton collisions produce
roughly equal numbers of $\pi^-$ and $\pi^+$). In the universe,
with strong magnetic field, such behavior can happen during
nucleus-nucleus collision. Maybe some of $\pi$ mesons, especially
$\pi^{-}$ mesons in the cosmic rays come from such nucleus-nucleus
collision, and they are accelerated (the acceleration mechanisms
may be that pions are attracted by proton collectivity and then
their trajectories are bent by magnetic field) by the electric
field produced by proton collectivity originating from
nucleus-nucleus collision in strong magnetic field.

In conclusion, we do a study of nuclear collision with strong
magnetic field. It is found that nuclear collision in strong
magnetic field (roughly $\sim10^{17}$ Tesla) causes the separation
of neutrons and protons from a nucleus. The formed proton
collectivity can produce electric field and accelerates proton or
charged pion meson up to very high speeds. Nuclear collision in
strong magnetic field can also produce asymmetric nuclear matter
transiently. The formed asymmetric nuclear matter may be used to
study isospin physics such as nuclear symmetry energy someday in
both heaven and terrestrial laboratory.

The work is supported by the National Natural Science Foundation
of China (10875151, 10740420550), the Knowledge Innovation Project
(KJCX2-EW-N01) of Chinese Academy of Sciences, the Major State
Basic Research Developing Program of China under No. 2007CB815004,
and the CAS/SAFEA International Partnership Program for Creative
Research Teams (CXTD-J2005-1).

\end{document}